\documentclass[runningheads]{llncs}
\usepackage[T1]{fontenc}
\usepackage{graphicx}
\usepackage{multirow}
\usepackage{algorithmic}

\usepackage[linesnumbered,ruled,vlined]{algorithm2e}
\usepackage{url}
\SetKwInput{KwInput}{Input}                
\SetKwInput{KwOutput}{Output}              

\usepackage{mathtools}
\usepackage{times}
\usepackage{epsfig}
\usepackage{xcolor}

\begin{document}
\title{SimBrainNet: Evaluating Brain Network Similarity for
Attention Disorders}
%
%
\author{Debashis Das Chakladar\and
Foteini Simistira Liwicki\and
Rajkumar Saini
}

\authorrunning{D Das Chakladar et al.}

\institute{Luleå University of Technology, 97187 Luleå, Sweden\\
\email{\{debashis.das.chakladar,foteini.liwicki,rajkumar.saini\}@ltu.se}
}
%
%
\maketitle              
\begin{abstract}
Electroencephalography (EEG)-based attention disorder research seeks to understand brain activity patterns associated with attention. Previous studies have mainly focused on identifying brain regions involved in cognitive processes or classifying Attention-Deficit Hyperactivity Disorder (ADHD) and control subjects. However, analyzing effective brain connectivity networks for specific attentional processes and comparing them has not been explored. Therefore, in this study, we propose multivariate transfer entropy-based connectivity networks for cognitive events and introduce a new similarity measure, ``SimBrainNet", to assess these networks. A high similarity score suggests similar brain dynamics during cognitive events, indicating less attention variability. Our experiment involves $12$ individuals with attention disorders ($7$ children and $5$ adolescents). Noteworthy that child participants exhibit lower similarity scores compared to adolescents, indicating greater changes in attention. We found strong connectivity patterns in the left pre-frontal cortex for adolescent individuals compared to the child. Our study highlights the changes in attention levels across various cognitive events, offering insights into the underlying cognitive mechanisms, brain dynamics, and potential deficits in individuals with this disorder.

\end{abstract}
\section{Introduction}
Attention disorder is a prevalent condition among both children and adults~\cite{slater2022can}. Most of the attention disorder studies focused on the diagnosis or classification of attention deficit hyperactivity disorder (ADHD) individuals. In a study using Near-infrared spectroscopy (NIRS), researchers compared typically developing (TD) children with those diagnosed with ADHD in different age groups. They found that children with ADHD showed reduced activation in the right and middle parts of the prefrontal cortex compared to TD children~\cite{yasumura2019age}. A similar study of resting state Functional magnetic resonance imaging (fMRI) of ADHD subjects with different age groups and TD subjects found that the middle temporal gyrus was the most significant region to discriminate ADHD and TD subjects~\cite{hong2017age}. In an fMRI-based neurofeedback study, authors found that the left inferior frontal gyrus and the left insular gyrus regions have significant interaction effects in the different age groups of ADHD subjects~\cite{criaud2020increased}. Electroencephalography (EEG)--based ADHD study highlights that absolute theta/beta band power ratio can be used as \textit{inattention index}~\cite{markovska2017quantitative} for ADHD subjects. It has been observed children with less attention compared to adults due to the high theta activity~\cite{clarke2019eeg}. Apart from identifying the brain regions, it is important to identify the underlying brain connectivity among those regions. Granger causality (GC)-based brain connectivity network is used for identifying workload levels during EEG-based mental arithmetic task~\cite{chakladar2021eeg}. However, due to the non-linear characteristics of EEG, GC is not a good choice for identifying information flow between brain regions~\cite{vicente2011transfer}. 
To overcome the issue, Transfer entropy (TE) assesses dynamic directional information flow between time series data in a non-linear manner \cite{schreiber2000measuring,gao2018electroencephalogram}. 
The Multivariate transfer entropy (MTE) can model better non-linear causal interaction between different brain regions than TE in EEG-based schizophrenia data analysis~\cite {harmah2020measuring}. The authors found a strong activation in the temporal lobe for schizophrenia patients. In many neuroscience studies, it's crucial to recognize the differences or similarities between healthy and pathological brain connectivity networks or between two different conditions while performing a cognitive task~\cite{mheich2020brain}. 
In~\cite{osmanliouglu2018graph}, authors developed graph matching and nodal features-based similarity measures for brain networks. In the dataset of traumatic brain injuries, they showed that patients had significantly lower matching accuracy compared to the control group. A similarity measure called "SimiNet" was created using node and edge properties with spatial information of node location~\cite{mheich2017siminet}. "SimiNet" performs well for identifying spatial differences in brain networks during a naming task involving animals and tools.
\par
\begin{figure}[!tb]
\centering
\includegraphics[scale=0.38]{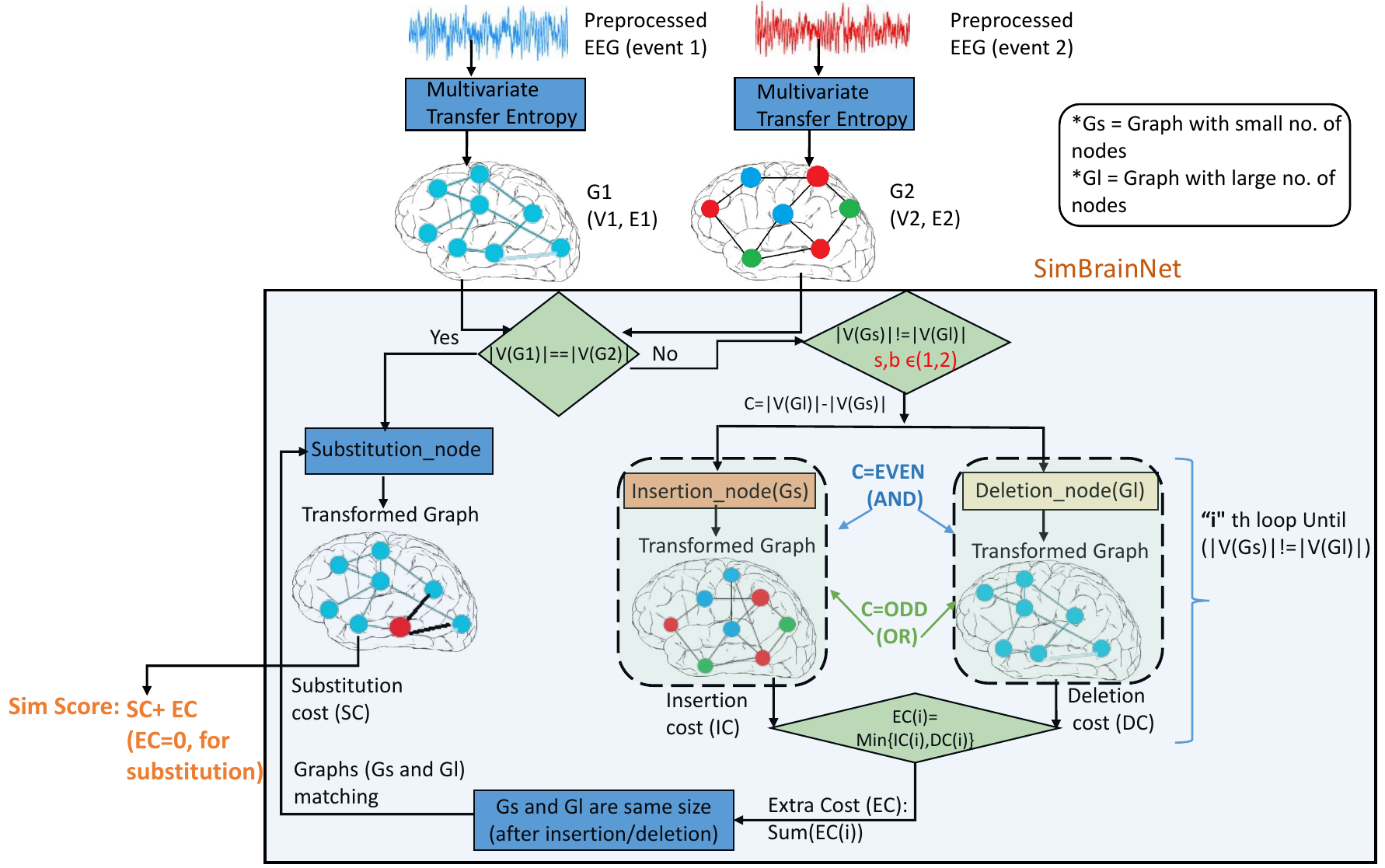}

\caption{
The proposed SimBrainNet model computes similarity scores between MTE-based brain connectivity networks (G1 and G2) using three algorithms: Substitution\_node, Insertion\_node, and Deletion\_node. If the number of nodes of G1 and G2 is equal, it calculates the substitution cost (SC); otherwise, it incurs extra cost (EC) through Insertion\_node, Deletion\_node, or both. The final score combines SC and EC.
}

\label{fig:1}
\end{figure}

However, most of the existing attention disorder studies~\cite{yasumura2019age,hong2017age} focused only on the activated brain regions for different age groups or perform ADHD classification using deep learning techniques~\cite{dong2020spatiotemporal,dubreuil2020deep}. 
Moreover, the existing similarity methods~\cite{mheich2017siminet,osmanliouglu2018graph,jie2018sub} of brain networks did not emphasize how the similarity score is related to human behavior and underlying brain connections. Thus, this paper highlights two novel approaches: (a) constructing MTE-based effective connectivity brain networks for two cognitive events, and (b) building a robust similarity measure (SimBrainNet) that computes the similarity between those brain networks (mentioned in (a)) using spatial neighbors of EEG channels. The framework of the proposed model is shown in Fig.~\ref{fig:1}. A detailed discussion of the framework is mentioned in Section 2.

\section{Methods}
\subsection{Dataset and Preprocessing}
In this research, we used a public EEG dataset~\cite{langer2017resource} that includes data from both developing brains and individuals with clinical disorders. We analyzed EEG data of 12 participants with attention disorders, divided into child ($G_1$: 6-10 years) and adolescent ($G_2$: 11-17 years) age groups, while they performed the 'Surround suppression' task. 
EEG recordings were captured using a 128-channel EEG Geodesic Hydrocel system at a sampling rate of 500 Hz. Based on the existing ADHD studies~\cite{rubia2018cognitive,konrad2010adhd,cao2023machine}, we found the relevant brain regions of ADHD. We identified the Brodmann areas (BAs) of those regions and selected $13$ EEG channels based on those BAs\footnote{\url{https://tinyurl.com/4fdszp7x}}.
The mapping of brain regions/cortexes to corresponding EEG channels is shown in Table~\ref{tab1}. Next, the raw EEG is preprocessed using a 1-40 Hz passband filter to eliminate high-frequency noise, and then we applied Independent Component Analysis (ICA) to remove EEG artifacts. We have removed the eye and muscle movements from EEG using ICA. Pz is used as a reference channel to extract the Independent Component.

\begin{table}[!htbp]
\centering
\caption{Mapping brain regions and EEG channels, where L/R denote the left and right hemispheres}
\begin{tabular}{|p{3.5cm}|p{6cm}|}
\hline
\textbf{Brain Regions} &  \textbf{EEG Channels (Brodmann Area)}\\
\hline
Medial prefrontal cortex & Fp1(10L),Fp2(10R),F1(24L),F2(24R) AFz(32)\\
Posterior cingulate cortex & Pz (23)\\
Inferior parietal cortex & P5(39L), P6(39R),CP3(40L), CP4(40R)\\
Occipital cortex & O1(17L), O2(17R), Oz (18)\\
\hline
\end{tabular}
\label{tab1}
\end{table}
\subsection{Construction of Brain Network}
The proposed MTE-based Brain connectivity network is constructed based on $13$ EEG channels (mentioned in Table~\ref{tab1}). For time-series data, $P$, $Q$, and $R$, we aim to find the information flow from source ($P$) to target ($Q$) system conditioned on $R$. Therefore, we denote $X$, $Y$, and $Z$ as the stationary stochastic processes describing the state visited by the systems $P$, $Q$, and $R$ over time. We consider the $X_n$, $Y_n$, and $Z_n$ as the stochastic variables obtained after sampling the processes at present $n$. The vector variables of past processes of $X$, $Y$ and $Z$ are denoted as $X_n^-=[X_{n-1}, X_{n-2},....]$, $Y_n^-=[Y_{n-1}, Y_{n-2},....]$ and $Z_n^-=[Z_{n-1}, Z_{n-2},....]$ respectively. Then, the MTE from $X$ to $Y$ conditioned on $Z$ is defined as follows~(\ref{eq:3})~\cite{montalto2014mute,chakladar2024brain}:
\begin{equation}
\begin{split}
MTE(X \rightarrow Y|Z) =\sum p(Y_n, Y_n^-, X_n^-, Z_n^-)
\log \Big(\frac{p(Y_n|Y_n^-,X_n^-,Z_n^-)}{p(Y_n|Y_n^-,Z_n^-)}\Big)
    \label{eq:3}
\end{split}
\end{equation}
Where $p(Y_n)$ is the probability of variable $Y_n$, and $p(Y_n|Y_n^-$) is the probability of observing $Y_n$ knowing the values of a past process $Y_n^-$.

\subsection{Finding Similarity in Brain Networks}
\label{sec:simnet}
The proposed Algorithm \textit{SimBrainNet} (Algorithm 1) consists of three sub-algorithms: \textit{Substitution\_node}, \textit{Deletion\_node} and \textit{Insertion\_node}.\\ 
\textbf{1. SimBrainNet}: 
This algorithm compares two MTE brain networks by transforming the first network ($G1(V, E)$) into the second one ($G2(V,E)$). In all four Algorithms, $V$ and $E$ are represented as the vertices and edges of the graph $G$. It maps EEG channels to respective brain lobes in a dictionary called $reg$ and computes their spatial neighbors using \textit{FieldTrip} software~\cite{oostenveld2011fieldtrip}, storing in another dictionary $ngh$. It calculates the $Sub\_cost$ using Algorithm 2 when the number of nodes is equal. For even disparities, it executes both \textit{Deletion\_node} and \textit{Insertion\_node} Algorithms to find the $Extra\_cost$, integrating it into $EC$. For odd disparities, it alternates between these Algorithms, computing the minimum cost in $EC$. For insertion, the parameter value ($p$ value from $0.01$ to $0.05$) is increased during the MTE experiment to include more significant nodes. Those nodes and their edge information are stored in a dictionary called $D\_ins$, passed during calling \textit{Insertion\_node}. After transformation, if the number of nodes is equal, the \textit{Substitution\_node} is called to find the matching between transformed graphs. Finally, the similarity score is obtained by combining the $Sub\_cost$ and $EC$.\\
\textbf{2. Substituion\_node:} 
In this Algorithm, we look for unique nodes in the first network ($G1$) compared to $G2$. For each such node ($n$), if there exists any node in $G2$ that is either in the same brain region or spatial neighbor of $n$, we mark the edges of a neighboring node ($p$) in $G2$ as replaced edges. In another case, we find the node ($imp\_node$) in $G2$ with the highest degree and use it as the substitute node. We calculate the substitution cost as the absolute difference between the edge values of the replaced edges and the unique node. This process is repeated for all unique nodes, and the total substitution cost is added to obtain the final cost for transforming the graph.\\
\textbf{3. Deletion\_node:} 
This Algorithm operates on the large graph $G2$, and a spatial neighbor dictionary ($D1$). We add edge value of neighbor nodes into a dictionary $k\_edgelist$. Then, it removes nodes in two ways: first, by identifying the neighbor node ($nbr\_node$) with the lowest edge value and marking it as the deletion cost ($del\_cost$), then adding the deleted node's edges to the neighbor node's list. If no neighbor is found, it removes the node with the lowest degree (for minimum information flow) in $G2$, along with its edges in the new graph ($G\_Dtrans$), labeling its edge values as deletion cost.\\
\textbf{4. Insertion\_node:} 
This Algorithm adds significant nodes to the small graph ($G1$). The $Di$ is constructed in such a way that the significant nodes are connected to any nodes of $G2$. We add the total edge values of such nodes to compute the insertion cost ($ins\_cost$). The transformed graph ($G\_Itrans$) is created after the insertion of the node.
\begin{algorithm}[!htb]
\caption{SimBrainNet}
\KwIn{MTE Brain networks $G1(V,E)$ and $G2(V,E)$}
\KwOut{$Sim\_score$}

\SetKwFunction{substitutionnode}{Substitution\_node}
\SetKwFunction{insertionnode}{Insertion\_node}
\SetKwFunction{deletionnode}{Deletion\_node}
$EC \gets 0$\\
$Dictionary$\hspace{0.05cm} reg $\gets\{\{F2,Fp2,AFz,F1,Fp1\} \text{:Frontal}, \{CP3, CP4, P5, P6, Pz\} :Parietal,\{O1,O2,Oz\} :Occipital\}$\\
$Dictionary$ \hspace{0.05cm} ngh $\gets \{F2\text{:\{AFz\}}, Fp2\text{:\{Fp1\}}, AFz\text{:\{F1,F2,Fp1,Fp2\}}, F1: \{AFz\}, Fp1\text{:\{Fp2\}}, CP3: \{P5\}, CP4: \{P6\}, P5: \{CP3\}, P6: \{CP4\}, Pz: \{Oz\}, O1: \{ Oz\}, O2: \{Oz\}, Oz: \{O1, O2, Pz\}\}$\\
$Dictionary \hspace{0.2cm} D\_ins \gets \{Sig\_channels: \{Edges; Edge values\}\}$\\
$sig\_nodes \gets$ \text{Significant nodes from MTE test} \text{(with increased $p$ value)}\\
$D\_ins= [sig\_nodes, \{sig\_nodes.edge, sig\_nodes.edge\_value\}$]\\    

\If{$(V(G1) = V(G2))$}
{
   $Sub\_cost \gets$ \substitutionnode($G1$, $G2$,$reg$,$ngh$)\\
}
\Else 
{
    $Gs\rightarrow \text{Graph with small no. of nodes between G1 and G2}$\\
    $Gl\rightarrow \text{Graph with large no. of nodes, between G1 and G2}$\\
    $i \leftarrow 1$\\
    \While{$((c \leftarrow \lvert V(Gl) - V(Gs) \rvert) \neq 0)$}
    {
     \If{$c$ is even}
      {
        $Ins\_cost[i] \leftarrow$ \insertionnode{$Gs, D\_ins$}\\
        $Del\_cost[i] \leftarrow$\deletionnode{$Gl, ngh$}\\
        $Extra\_cost[i].add(\min(ins\_cost[i], del\_cost[i]))$\\
      }
      $EC \leftarrow \sum_{i=1}^{c} Extra\_cost[i]$\\
      \Else{
        \For{$j \leftarrow 1$ \KwTo $c$}
        {
            $Ins\_cost[j] \leftarrow$ \insertionnode{$Gs, D\_ins$}\\
            $Tot\_inscost \leftarrow Tot\_inscost + Ins\_cost[j]$\\
        }
        \For{$k \leftarrow 1$ \KwTo $c$}
        {
            $Del\_cost[k] \leftarrow$ \deletionnode{$Gl, ngh$}\\
            $Tot\_delcost \leftarrow Tot\_delcost + Del\_cost[k]$\\
        }
        $EC \leftarrow \min(Tot\_inscost, Tot\_delcost)$\\
        }
        $i \leftarrow i + 1$\\
    }
 }
\text{Call lines (7-8)}\\
$Sim\_score \gets Sub\_cost + EC$\\
\KwRet{$Sim\_score$}\\
\end{algorithm}

\begin{algorithm}[!htbp]
\caption{Substituion\_node}
\SetAlgoLined
\KwIn{MTE Brain networks $G1$, $G2$, dictionaries $D$ and $D1$}
\KwOut{$subst\_cost$}

$uncom\_nodes \gets V(G1) - (V(G1) \cap V(G2)$)\\
$subst\_cost \gets 0$\\

\ForEach{(($n \in uncom\_nodes) \And (k \in V(G2))$}
{
    \If{any($D[n]==D[k])\lor (n==any(D1[k])$)}
    {
         $p \gets any (k), \text{where } k \in D[n]$\\
         $G2\_rep\_edges \gets \sum_{\text{edge}\in G2.\text{edges}(p)} \text{G2.\_MTE\_val}(\text{edge})$ \\
    }
    \Else
    {
        $imp\_node \gets Max(degree($k$), k\in V(G2))$ \\    
        $G2\_rep\_edges \gets \sum_{\text{edge}\in G2.\text{edges}(imp\_node)} \text{G2.\_MTE\_val}(\text{edge})$ \\   
    }
    $G1\_old\_edges \gets \sum_{\text{edge} \in G1.\text{edges}(n)} \text{G1.\_MTE\_val}(\text{edge})$\\
    $subst\_cost+= |\text{G2\_rep\_edges.value} - \text{G1\_old\_edges.value}|$\\
}
$G\_Strans \gets G11 \{(V1 \in p\cup imp\_node),(E1 \in G2\_rep\_edges)$\}\\

\KwRet{$subst\_cost$}
\end{algorithm}

\begin{algorithm}[!htbp]
\caption{Deletion\_node}
\SetAlgoLined
\KwIn{MTE Brain network $G2=(V,E)$, dictionary $D1$}
\KwOut{$del\_cost$}
$\text{Dictionary } k\_edgelist:[{node, edge\_val}] \gets NULL$\\

\For{($n \in V(G2$))}
{
 \If{($D1[n] \in V(G2)$)}
 {
    \ForEach{($k \in D1[n], Edge(n,k)\neq NULL$)}
    {
       $edge\_val (k)\gets \sum_{\text{edge}\in G2.\text{edges}(k)} \text{G2.\_MTE\_val}(\text{edge})$ \\
       $k\_edgelist.add(k, edge\_val (k))$\\     
    }
    $Min\_edge \gets Min(k\_edgelist[edge\_val])$\\
    $del\_cost \gets Min\_edge.value$\\ 
    $nbr\_node \gets k\_edgelist[Min\_edge]$ \\
    $G\_Dtrans \gets G(V - \{n\}, \{edges(nbr\_node) \cup edges(n)\})$\\
}
 \Else
 {
    $d\_node \gets Min(degree($p$), p\in V(G2))$ \\    
    $G2\_del\_edges \gets \sum_{\text{edge}\in G2.\text{edges}(d\_node)} \text{G2.\_MTE\_val}(\text{edge})$ \\   
    $del\_cost \gets G2\_del\_edges.value$\\
    $G\_Dtrans \gets G(V - \{d\_node\}, \{E - G2\_del\_edges\})$\;  
 }
}
\KwRet{$del\_cost$}\\
\end{algorithm}

\begin{algorithm}[!h]
\caption{Insertion\_node}
\SetAlgoLined
\KwIn{\text{MTE Brain network} $G1(V,E), \text{Dictionary } Di$}
\KwOut{$ins\_cost$}


\ForEach{(($n \in Di[sig\_nodes]) \And (k \in V(G1)))$ }
{
        \If{$(\exists Di[n.edge] \in (n, k))$}
        {
        $edge\_val (n)\gets \sum_{\text{edge}\in Di[n.edge]} \text{Di[n.edge.value]}(\text{edge})$ \\
        $ins\_cost \gets edge\_val(n)$\\
        }
}


$G\_Itrans \gets G1 \{(V \cup Di[sig\_nodes]), (E \cup Di[sig\_nodes.edge])$\}\\
\KwRet{$ins\_cost$}\\
\end{algorithm}

\section{Results}

\subsection{Brain Connectivity Network analysis}
\label{ref:brain}
\begin{figure}[!tb]
\centering
\includegraphics[scale=0.42]{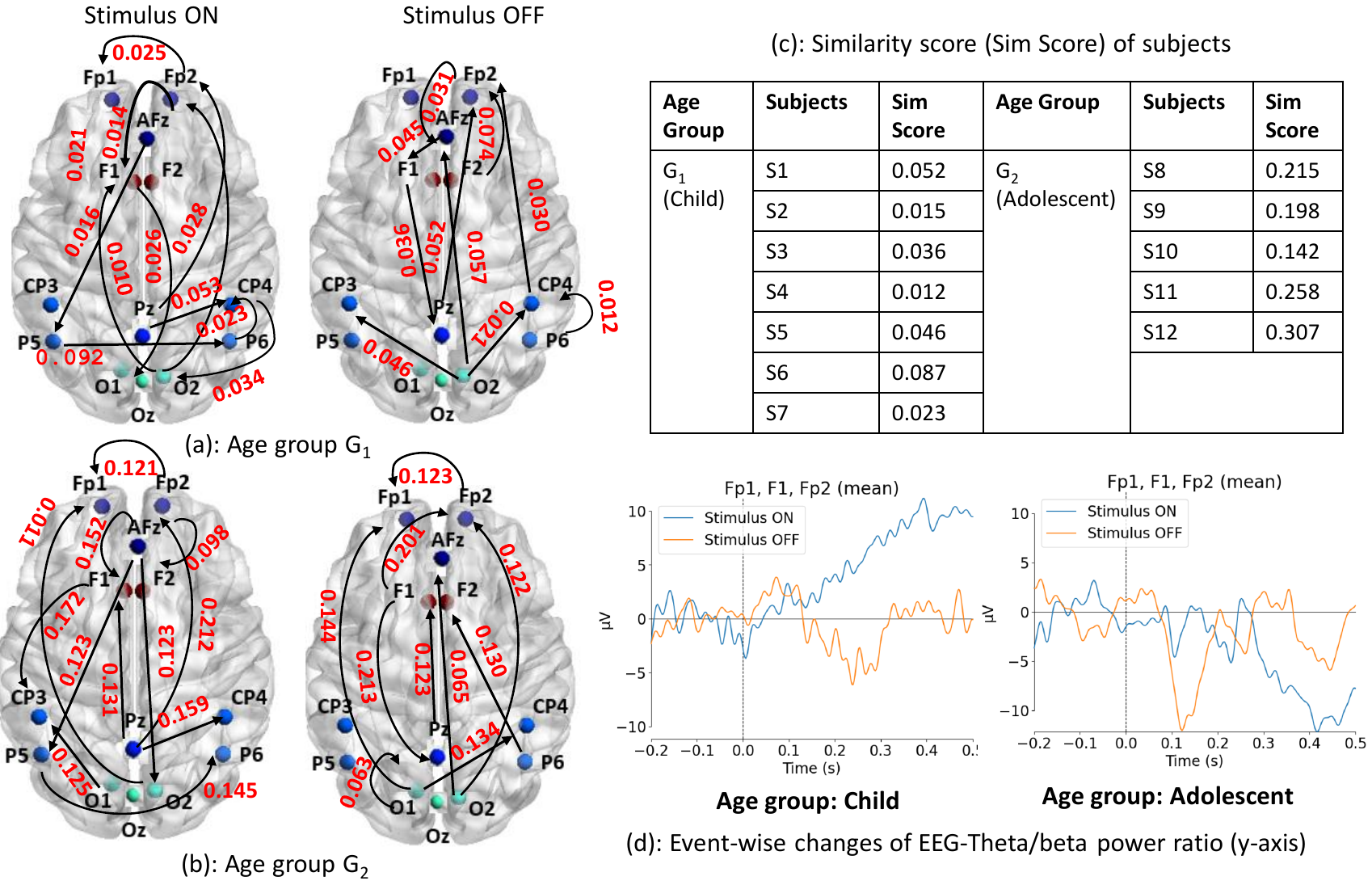}
\caption{Age group-wise (a,b) common connectivity patterns of brain networks. Different colored nodes represent different brain regions whereas the average MTE value between the source and target node for all subjects represents the directed edge value. (c) The similarity score/Sim Score of all subjects based on event-wise MTE brain networks. (d) EEG-Theta/Beta Ratio (TBR), referred to as \textit{inattention index}~\cite{markovska2017quantitative} for age groups.
}
\label{fig:2}
\end{figure}
This section discusses the brain connectivity network using MTE. The connectivity network is created for two events: stimulus ON (event 1)/stimulus OFF(event 2) of the surround suppression task. The average connectivity result of the two age groups is shown in Fig.~\ref{fig:2}(a-b). First, we find the event-wise common brain activation irrespective of age group, next, we focus on age group-wise connections. For event 1, a common connectivity pattern is found between parietal and frontal brain regions (Pz$\rightarrow$ Fp2, AFz$\rightarrow$P5), and occipital to frontal brain regions (O2$\rightarrow$Fp1/Fp2) for both age groups. On the other hand, occipital to central-parietal (O2$\rightarrow$CP3/CP4, O1$\rightarrow$ CP4) and frontal to parietal (F1$\rightarrow$Pz) regions connectivity patterns are found for event 2. Strong connectivity patterns are observed in the left pre-frontal cortex (PFC) for age group $G_2$ compared to $G_1$, with more MTE connections in EEG channels Fp1 and F1. This finding aligns with a similar ADHD study where younger ADHD subjects exhibit greater left PFC activation than ADHD children~\cite{yasumura2019age}.
An independent samples $t$-test was used to compare the average MTE connectivities between two age groups: $G_1$ and $G_2$. We set the event-wise MTE connectivities of each subject in both age groups as the dependent variable, while the age group is considered the independent variable. Descriptive statistics showed that the mean MTE connectivity is $0.023$ (SD = $0.206$) for $G_1$ and $0.132$ (SD = $0.059$) for $G_2$. The independent samples $t$-test indicated a significant difference present in MTE connectivities between the two age groups, $t(229)$ = $6.155$, $p =0.001$.

\subsection{Analysis of SimBrainNet}
\label{ref:sim}
In this section, we assess the similarity score between MTE-based brain connectivity networks of two events during the surround suppression task. The results, depicted in Fig.~\ref{fig:2}(c), indicate that subjects in $G_1$ exhibit lower similarity scores, suggesting distinct attention graphs for different events, reflecting more changes in attention with underlying brain dynamics. Conversely, subjects in $G_2$ show higher similarity scores, indicating minimal changes in attention between events. The variations in attention among ADHD subjects can be further understood by analyzing the Theta/beta ratio (TBR) of EEG signals, a metric which is useful for identifying \textit{inattention} in children and adolescents with ADHD~\cite{markovska2017quantitative}. Here, we average event-wise all subject EEG data for two age groups and calculate TBR for more activated channels (Fp1, F1, Fp2) in MTE networks. For both events, we noticed higher TBR values in children with ADHD, suggesting greater inattention compared to adolescents with ADHD (refer to Fig.~\ref{fig:2}(d)). This supports our observations on how similarity scores relate to attention changes.
Computational complexity of SimBrainNet is $\mathcal{O}(n.m.e)$, where $n$, $m$, and $e$ are the nodes in $G1$, $G2$, and the edges in the graphs. 
Creating MTE graphs takes more time because of the high-dimensional EEG data. 
Removing non-significant connections produces lower-dimensional MTE graphs, which keep SimBrainNet’s complexity manageable. Here, each MTE graph creation takes $8$-$10$ minutes with an $8$ GB GPU and $32$ GB RAM. However, we anticipate decreased graph creation time with more powerful hardware.

\subsection{Performance analysis}
\label{ref:com}
For generalization, the proposed model is tested (refer to Table~\ref{tab:2}) using two versions (Release $9$ and Release $10$) of the CMI-HBN dataset~\cite{alexander2017open}. We randomly selected $15$ subjects with attention disorder from each version based on age groups of $G_1$ and $G_2$. To ensure consistency, we computed the similarity score for the same cognitive task with similar events, averaging the scores across age groups. For these datasets, we achieved similar findings for similarity scores across age groups like the experimental dataset.
\begin{table}[!htbp]
\caption{Performance analysis based on other EEG-based attention disorder datasets}
\begin{center}
\begin{tabular}{|p{3.7cm}|p{3cm}|p{2cm}|}
\hline
    \textbf{{Dataset}} & \textbf{{Sub (Group)}} & \textbf{{Avg SimScore}}\\
    \hline
    \multirow{3}{*}{CMI-HBN (Release 9)\cite{alexander2017open}} & {S1-S8 ($G_1$)} & {0.035}\\
    \cline{2-3}
    & {S9-S15 ($G_2$)} & {0.153}\\
     \hline
    \multirow{3}{*}{CMI-HBN (Release 10)\cite{alexander2017open}} & {S1-S6 ($G_1$)} & {0.055}\\
    \cline{2-3}
    & {S7-S15 ($G_2$)} & {0.172}\\
    \hline
\end{tabular}
\label{tab:2}
\end{center}
\end{table}
\section{Conclusion}
This paper proposed a similarity method called ``SimBrainNet" to find similarities between two brain networks of attention disorder subjects with two age groups (child and adolescent). We noticed that the similarity score is high for the adolescent group compared to the child group. 
The similarity score obtained from SimBrainNet can improve ADHD diagnosis and create personalized treatment plans. It can also be helpful for developing neurofeedback protocols and cognitive training programs. In this study, we experimented with a small number (i.e. $12$) of ADHD subjects. Therefore, in the near future, we will evaluate our proposed model using ADHD datasets with more number of subjects. 

\begin{credits}
\subsubsection{\discintname}
The authors have no competing interests in the paper as required by the publisher.
\end{credits}
%
%
%
\bibliographystyle{splncs04}
\bibliography{Paper-2277}
\end{document}